\documentclass[conference,a4paper,10pt]{IEEEtran}
 \usepackage{mathtools}
 \usepackage{amsfonts}
 \usepackage{amsmath}
 \usepackage{amsthm}
 \usepackage{amssymb}

\usepackage{tablefootnote}
\usepackage{url}

\usepackage{algorithm}
\usepackage{algorithmicx}
\usepackage{algpseudocode}
\usepackage{booktabs}
\usepackage{textcomp}
\usepackage{enumitem}
\usepackage{subfigure}
\usepackage{graphicx,epstopdf}
\usepackage{xcolor}
\usepackage{url}
\usepackage{cite}
\usepackage{soul}
\usepackage{balance}
\usepackage{multirow}

\definecolor{hgreen}{rgb}{0, 0.5, 0}

\ifCLASSINFOpdf
\else
\fi
\begin{document}

%
\title{Electric Field Evaluation of Reconfigurable Intelligent Surface in Wireless Networks}
\author{Zhuangzhuang~Cui, Franco Minucci, Rizqi Hersyandika, Rodney Martinez Alonso,\\ Andrea P. Guevara, Hazem Sallouha, and Sofie Pollin\\
WaveCoRE, Department of Electrical Engineering (ESAT), KU Leuven, Belgium\\
Email: \texttt{\{firstname.lastname\}@kuleuven.be}}

%



\maketitle

\begin{abstract}
Reconfigurable intelligent surface (RIS) used as infrastructure in wireless networks has been a trend, thanks to its low cost and high flexibility. Working in many ways including reflective mirrors and phase-shifted surfaces, RIS is able to enhance the coverage in communications and provide more degrees of freedom for sensing. However, the key issue lies in how to place RIS in accordance with the regulations for electromagnetic field (EMF) exposure, which requires refined evaluations. In this paper, we first investigate the regulations in terms of E-field. Then, relevant deployment characteristics are evaluated jointly: the minimum distance from the base station (BS) to the RIS, and the minimum height of the RIS are given for a given BS power limit and as a function of the number of RIS elements. The ray-tracing simulations verify the correctness of our analysis. Besides, different frequency ranges (FRs) and radiation patterns of RIS elements are investigated. The results show that the EMF exposure risk is negligible when RIS works in the reflective-only (RO) mode. However, when it works in the beamforming (BO) mode, its placement should be well specified based on our analytical framework to comply with the regulations of E-field limit in general public scenarios. Finally, we provide an E-field measurement methodology and low-cost solutions in terms of general wireless networks and 5G standalone networks, which pave the way for real-world evaluation in future work.

\end{abstract}

\begin{IEEEkeywords}
6G, electric field, reconfigurable intelligent surface, beamforming, power density, spectrum regulation.
\end{IEEEkeywords}

\section{Introduction}
Reconfigurable intelligent surface (RIS) technology has been widely acknowledged as one of the potential 6G solutions to enhance communications coverage and assist sensing in wireless networks, thanks to its capacity to control channels and a high degree of freedom (DoF) by employing a high amount of elements. As most existing studies focus on the benefits the RIS brings to wireless networks \cite{ris_tutorial}, a key question lies in whether it creates risks simultaneously in terms of electromagnetic field (EMF) exposure due to the coherent superposition of large numbers of elements.

The capillar diffusion of radio-transmitting devices started sparking questions, and research about possible health effects caused by the non-ionizing radiation \cite{balmori2014electrosmog}. However, research on the biological effect of electromagnetic (EM) waves on the human body shows that high-frequency fields are absorbed in the skin surface and only a small portion of energy penetrates the underlying tissues \cite{salovarda_malaric}. Since the RIS can work in any frequency band, it becomes crucial to comprehensively investigate its impact on the electric field (E-field) before we deploy it in real networks. Moreover, to comply with existing regulations that are diverse in terms of frequencies, countries, and organizations (such as WHO, ITU, and ICNIRP), one urgently requires a complete analytical framework to evaluate the E-field from the RIS side, following previous evaluations of multi-antenna systems \cite{mimo_sar1}. With the local regulation, the framework can provide placement insights of RIS, so that the EMF exposure risk can be understood and mitigated.

To evaluate the E-field in RIS-assisted networks, one way is to employ EM simulation software that can directly calculate EM field strength. The work in \cite{peng22} used CST Microwave Studio software to study the E-field of RIS in the near-field regions; however, the power attenuation from the base station (BS) to the RIS is not considered. This is needed as the BS is already complying with EM field regulation, and only in special scenarios with a very short RIS-BS distance possible problems occur. Moreover, the computational complexity is very high and the flexibility is low when using simulation software. A simplified EM wave simulation software, i.e., Wireless InSite, is used in this paper to verify our analytical results; however, simulations of an area of 10 m by 10 m took several days for a single deployment scenario of a relatively small $8\times 8$ RIS,  which suggests difficulties when a high amount of elements is considered. Simplified analytical models are hence more appropriate for bounding the zones for detailed simulation or eventually in-the-field measurements.

To obtain tractable E-field results, the work in \cite{marco_20} analyzed the E-field for three working modes categorized by the phase-shifting capacity, where the different scaling laws were observed as a function of distance. With E-field expressions, the received power is generally evaluated in most existing works \cite{emil21,jiang23,georgios22,meapl,placement21,zach_ris}. In \cite{emil21}, the authors investigated the normalized RIS gain and the near-field propagation, where the interplay between beamforming gain and far-field distance denoted as $d_{\rm F}$, is illustrated. Following that, the work in \cite{jiang23} considered not only the Fresnel region but also the reactive region in the near field, however, the focus lies on the impact on channel capacity. Besides, real-world measurements using RIS were conducted in \cite{georgios22,meapl}, where the authors tested received power performance when using the beamforming mode, and statistical path loss models were proposed. Only a few works studied the placement of RIS \cite{placement21,zach_ris}. The authors in \cite{placement21} studied the effective illumination on the RIS considering a narrow beam from the BS, whereas \cite{zach_ris} studied the impact of RIS geometries and locations on communication performance. It is found that there are very limited works that focus on the detailed E-field evaluation and provide the corresponding deployment insights.


In this paper, we first provide an overview of the regulation for EMF exposure, which also applies to RIS-assisted networks. Based on the analytical framework and simulations, the E-field is comprehensively evaluated to provide RIS deployment insights. In summary, the main contributions include the following aspects.
\begin{itemize}
    \item A complete analytical framework is built, considering two working modes: Beamforming mode (BO) and Reflective mode (RO).  This enables the deployment evaluations for arbitrary scenarios such as the power limit at the BS, as well as patterns and numbers of elements in the RIS.
    \item A comprehensive evaluation of the E-field is conducted, considering various sizes of RIS and different Frequency ranges (FRs) at RO and BO modes, which shows working in RO has very little influence on EMF exposure, while the limits of RIS height and BS-RIS distance considering the BO mode are given.
    \item A regulation overview and detailed measurement methodologies are provided, where we investigate the policies worldwide. Moreover, general measurement procedures are introduced for real-world evaluation and 5G networks.
\end{itemize}

The rest of the paper is organized as follows. Section II introduces the regulation of electric fields in wireless networks. In Section III, we introduce the system model and provide the analytical expressions of the E-field. Section IV first verifies the analysis by using EM simulation software, and then a comprehensive evaluation is provided, along with the insights for RIS deployment. In Section V, we introduce the measurement principles and special considerations for future real-world evaluations. We conclude the paper in Section VI.

\section{Regulation Overview}

Exposure guidelines are established
for the protection of humans exposed to E-fields ranging from 100 kHz to 300~GHz.
The International Commission on Non-Ionizing Radiation Protection (ICNIRP) issues comprehensive guidelines delineating permissible power levels across various frequency bands for radio-emitting sources \cite{icnirp_web}. Evaluation of a radio source's maximum allowable power involves the estimation of the Specific Absorption Rate (SAR) \cite{sar_mimo}, a metric quantifying the energy absorption within tissues resulting from exposure to high-frequency electromagnetic fields per unit mass.


The International Telecommunication Union (ITU) has developed a set of recommendation documents for human exposure limits and guidelines for the assessment of electromagnetic fields radiation from any given base stations in the ITU-T K.52 and K.100 recommendations~\cite{itu_k52,itu_k100}. The ITU recommends two different exposure limits for the occupational and general public zones in the frequency range 2-300~GHz~\cite{itu_k52,itu_k100}. The occupational exposure limits consider areas where individuals work occasionally and may be exposed to higher electromagnetic fields and radiation levels. The general public exposure limits consider the areas where the general population is located. These zones are further away from the RF sources than the occupational zones. The limits in this zone are generally lower than the occupational ones.

Within the frequency range from 100~kHz to 6~GHz, the ICNIRP defines the electromagnetic exposure limits considering a margin from the equivalent radiation causing an induced local thermal variation of 2 to 5 \textcelsius~on a 10~g tissue sample over a period of 6~minutes. For whole-body exposure, the thermal variation is limited to 1\textcelsius~\cite{emf_guideline}. Nevertheless, these general guidelines are the recommended maximum limits but each regulatory authority might define different limits. Focusing on microwave bands, Table~\ref{tab1} lists the updated E-field exposure limits from ITU, WHO, ICNIRP, and some selected countries for the range of frequency up to 300~GHz. Moreover, these limits are consecutive as a function of the operating frequency. It is noted that these E-field limits are computed for a reference field at 900 MHz. Moreover, the limits are cumulative, summing up all E-fields from all sources at a given location. 

\begin{table}[!t]
\caption{Regulation overview of E-field limits for general public considering various authorities.}
\centering
\begin{tabular}
{|l|c|c|}
\hline
\textbf{Authority} & \textbf{E-field limit (V/m)} & \textbf{Frequency range} \\ \hline
\multirow{2}*{ITU \cite{itu_k100}} & 27.5 -- 61.5 & 400 -- 2000 MHz \\
         ~           & 61 & 2 -- 300 GHz \\ \hline
WHO \cite{who} & 41.25\tablefootnote{The exposure limit differs across various countries, but the majority of them adopt 41.25 V/m without the frequency range specified in the reference.} &  N/A \\ \hline
\multirow{2}*{ICNIRP \cite{emf_guideline}} & 27.5 -- 61.5 & 400 -- 2000 MHz\\
  ~  & N/A\tablefootnote{In this frequency range, compliance is determined by the power density limit of 10~W/m$^2$, within the reactive near-field zone.} & 2 -- 300 GHz \\ \hline
\multirow{2}*{USA \cite{emf_usa}} & 27.46 -- 61.4 & 300 -- 1500 MHz  \\
        ~            & 61.4  & 1.5 -- 100 GHz  \\ \hline
\multirow{2}*{Flanders \cite{emf_belgium}}  & 13.7 -- 30.7 & 400 -- 2000 MHz  \\
   ~     & 30.7 & 2 -- 300 GHz  \\ \hline
\multirow{3}*{China \cite{emf_china}} & 12 & 30 -- 3000 MHz  \\
  ~    & 12 -- 27 & 3 -- 15 GHz \\
  ~    & 27 & 15 -- 300 GHz \\ \hline
\end{tabular}
\label{tab1}
\end{table}

In general, the ITU recommendation for assessing the EMF exposure levels is based on the time-averaged rate of energy transfer during the exposure time. Some extrapolation methods are recommended for OFDM signals allowing for evaluating the E-field level of a reference signal transmitted by the BS at a constant power level, and extrapolating this value to the maximum possible power over the whole band. This method has shown high accuracy for instance for exposure assessments for LTE operating under the Frequency Division Duplexing regime where the downlink and uplink are separated in the frequency domain~\cite{itu_k52,itu_k100}. However, new technological developments in duplexing, dynamic bandwidth aggregation, beamforming, and the use of RIS might be a challenge from the regulatory perspective to the control and exposure assessment. Besides, the ITU recommended the assessment domain boundary (ADB) for the measurement, which suggests that the squared evaluation area size $D$ is determined by the EIRP of the source, and the measured height can be calculated by $H_b=\max(D\tan(\alpha),3.5~m)$ considering a down-tilted angle $\alpha$~\cite{itu_k100}. It is seen that the measured height corresponds to the main lobe direction when considering a directional beam. Moreover, the domain of investigation can be restricted only to those points where the level of exposure is expected to be relevant or maximum, according to ITU-T K.70 \cite{itu_k70}.





\begin{figure}[!t]
  \centering
   {\includegraphics[width=0.9\linewidth]{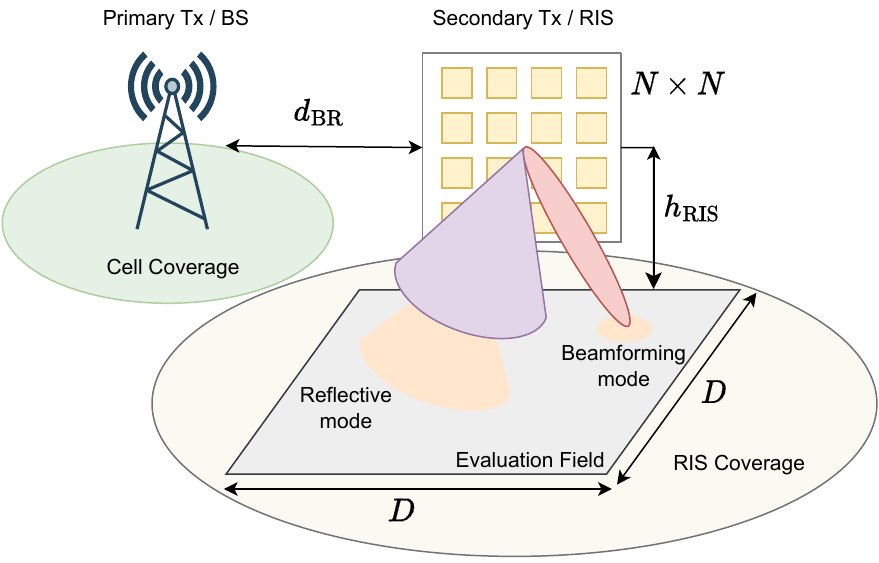}}
  \caption{System model.}
  \label{riscom}
 \end{figure}


\section{System Model}
Considering our system model in Fig.~\ref{riscom}, RIS is used as a \textit{secondary transmitter} that is a substitute for the primary transmitter, i.e., the BS. In this scenario, the RIS is able to generate its own coverage to top up the insufficient coverage of the cellular BS. Therefore, to evaluate the E-field within the RIS coverage, the additive influence of cell coverage is negligible. However, the distance between BS and RIS denoted as $d_{\rm BR}$ determines the received power at the RIS and afterward impacts the E-field. In addition, the E-field is influenced by the number of elements $N^2$ and the height of the RIS $h_{\rm RIS}$. In practice, for a given E-field limit in the evaluation area, these parameters can be adjusted and optimized, e.g., the height can be increased to reduce the E-field on the ground. It is noted that $D$ is the length of a squared evaluation area, which can be determined by the transmit power at the RIS. However, when we merely consider a peak E-field, we can use a large $D$, e.g., $10\times 10$~m, whose area is large enough to guarantee it containing all the peak values.

\subsection{Link from BS to RIS}
The BS is the transmit power source providing the power to the  RIS. Therefore, it is important to consider the path loss between BS and RIS, where we refer to the free-space path loss, expressed as
\begin{equation}
    FSPL=\left(\frac{\lambda}{4\pi d_{\rm BR}}\right)^2,
\end{equation}
where $\lambda$ is the wavelength of the carrier frequency.

As an example, the Effective Isotropic Radiated Power (EIRP) limit is 75~dBm considering 100~MHz bandwidth in FCC \cite{fcc_75}. However, considering different FRs in 5G New Radio (NR), this limit for given aggregate bandwidth can be 47~dBm or 59~dBm in FR1 and FR2~\cite{3gpp_fr1_power,3gpp_fr2_power}, respectively. Thus, we denote it as $P_{max}$. The received power at the RIS is calculated by
\begin{equation}
    P_{RIS} = P_{max} \cdot FSPL.
\end{equation}

\subsection{Link from RIS to User}
Considering a general squared RIS layout, there are $N\times N$ elements in total. For evaluation in the $D \times D$ area, we use a Cartesian coordinate system, where the central point of the RIS is $(0,0,h_{\rm RIS})$, and the user's height is $h_u$.

To express the relative geometric relationship between the user location and each element in the RIS, the first step is to obtain the coordinates of each element in the RIS, where there are two situations considering $N$ is even or odd. For the $n$th element in the RIS, the coordinate can be expressed as
\begin{equation}
    x_n=x_s+ p d_s,\;
    h_n=h_s + p d_s, \; \text{$N$ is odd},
\end{equation}
where $p\in\left(\lceil -\frac{N}{2} \rceil, \lfloor \frac{N}{2} \rfloor\right)$ and $y_n=0$. Considering $N$ is even, $x_n$ and $h_n$ can be expressed as
\begin{equation}
    x_n=x_s+ q \frac{d_s}{2},\;
    h_n=h_s + q \frac{d_s}{2}, \; \text{$N$ is even},
\end{equation}
where $q\in\left(-N+1, N+1 \right)$ with the length of $\rm{dim}(p)-1$.
For the coordinate of user $(x_u, y_u, h_u)$, the 3D distance between the user and $n$th element can be expressed as
\begin{equation}
    r_n=\sqrt{(x_n-x_u)^2+(y_n-y_u)^2+(h_n-h_u)^2}.
\end{equation}
Moreover, the azimuth and elevation angles are calculated by
\begin{equation}
    \begin{aligned}    \psi_n&=\arctan\left(\frac{x_n-x_u}{y_n-y_u}\right),\\
\theta_n&=\arctan\left(\frac{h_n-h_u}{\sqrt{(x_n-x_u)^2+(y_n-y_u)^2}}\right).
    \end{aligned}
\end{equation}

For the practical gain pattern of the RIS elements, we mainly use $G_n=\cos(\theta_n)^3$ for $\theta_n$ and $\psi_n$ \cite{meapl}. However, different patterns may be used to evaluate their impacts.

With the geometric representation, we can first formulate the received power at the user, which is expressed as
\begin{equation}
P_u=P_{RIS}\left|\sum_{n=1}^{N^2}\frac{\sqrt{G_n} e^{-j(\phi_n-\phi_d)}}{r_n}\right|^2,
\label{eq:efield_user}
\end{equation}
where $\phi_n = \frac{2\pi}{\lambda}r_n$ is the applied phase and $\phi_d$ is the phase at the desired location of $(x_d,y_d,h_d)$. Thus, the situations of $\phi_d=0$ and $\phi_n-\phi_d=0$ correspond to the reflective only (RO) mode and beam spotting (BO) modes (perfect phase shift), respectively.


Finally, the E-field can be calculated by \cite{vittorio}
\begin{equation}
\label{eq:ele_field_f}
    E_u = \sqrt{\frac{\eta Pu}{2\pi}},
\end{equation}
where $\eta$ is the wave impedance in free space which is $120\pi$.

 \begin{figure}[!t]
  \centering
   \subfigure[Wireless Insite simulation result]{\includegraphics[width=3in]{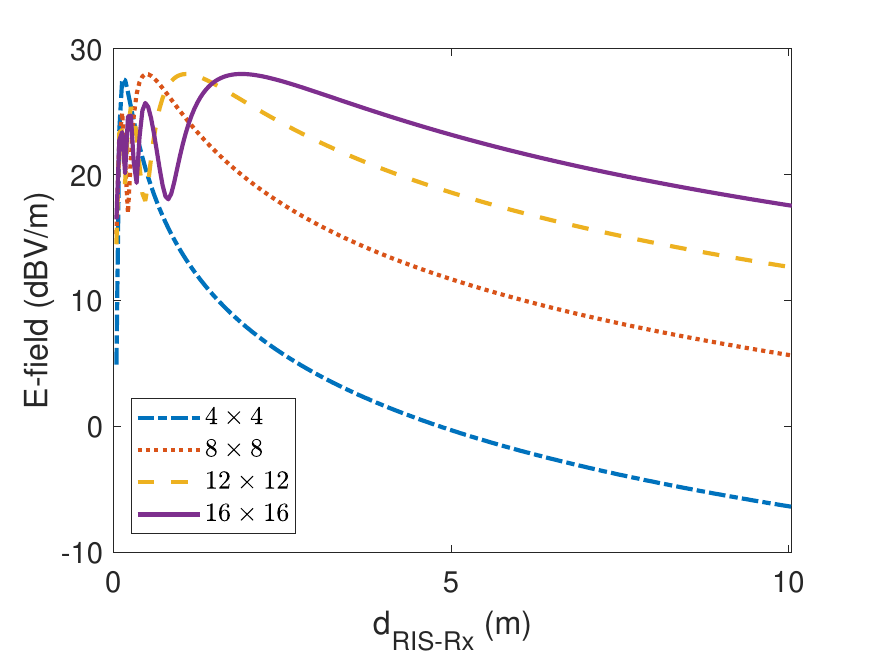}}
   \subfigure[Analytical result ]{\includegraphics[width=3in]{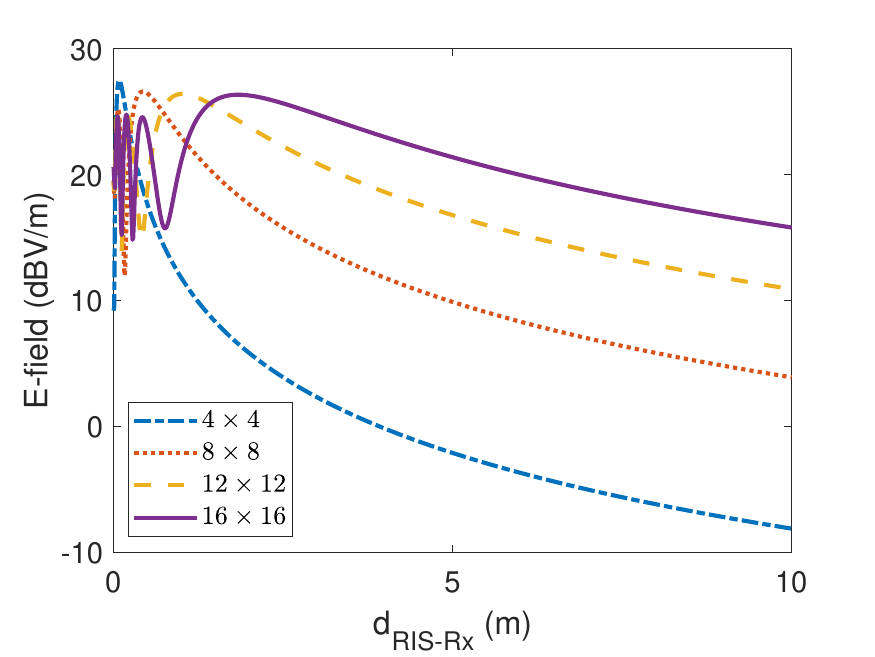}}
  \caption{Comparison of E-field considering various RIS sizes with omnidirectional element and 1~mW power at the RIS.}
  \label{e-field-comparison}
 \end{figure}
 \begin{figure*}[!t]
  \centering
   \subfigure[Wireless InSite]{\includegraphics[width=1.75in]{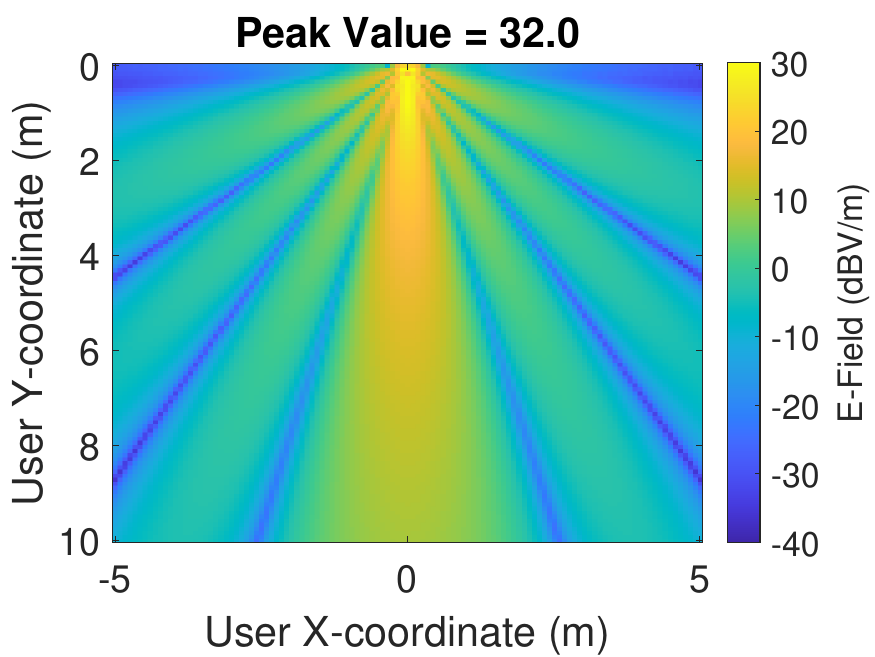}}
   \subfigure[$h_{\rm RIS}=1.5$ m, RO]{\includegraphics[width=1.75in]{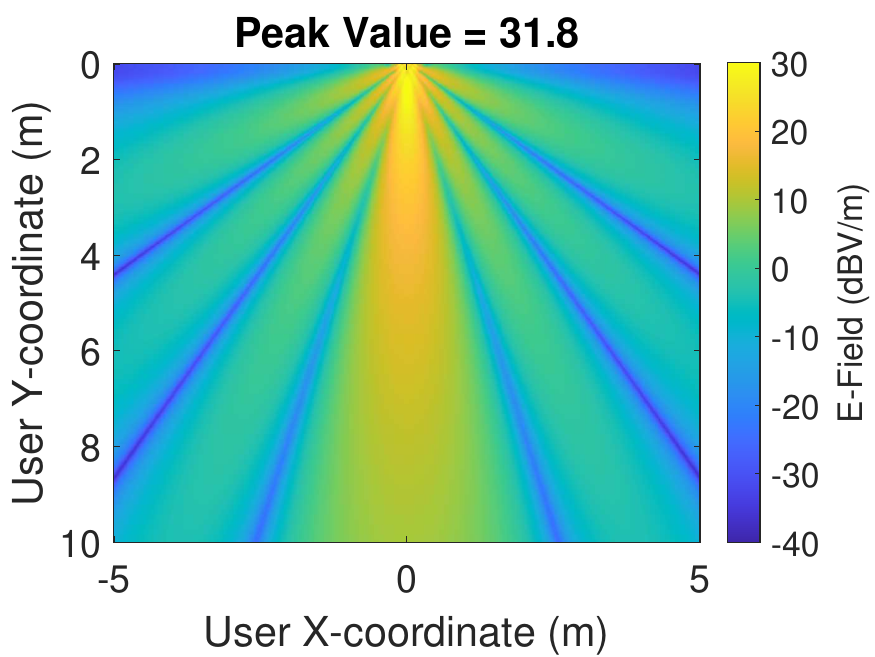}}
   \subfigure[$h_{\rm RIS}=3$ m, RO]{\includegraphics[width=1.75in]{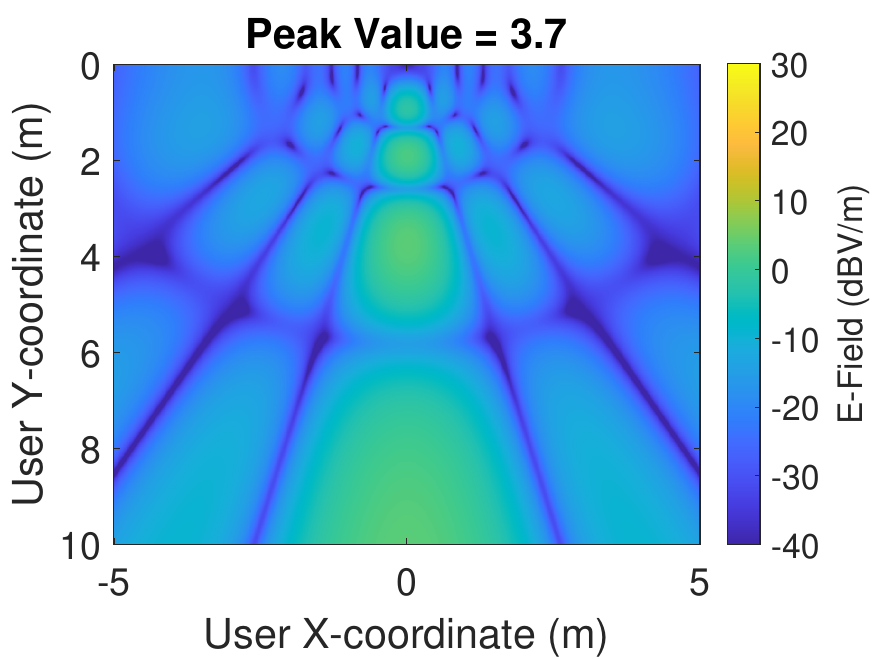}}
   \subfigure[$h_{\rm RIS}=3$ m. BO]{\includegraphics[width=1.75in]{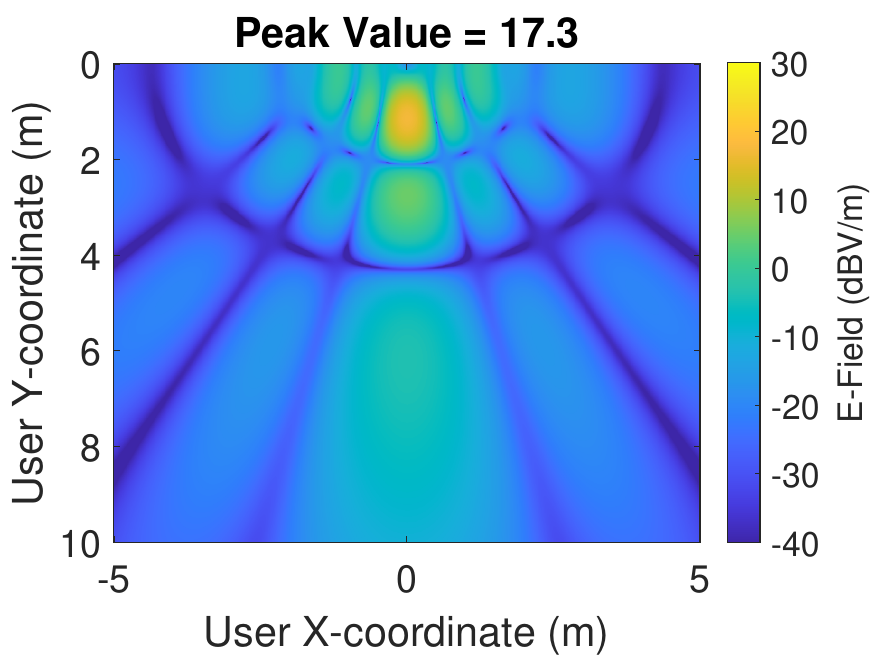}}
  \caption{E-field map in a $10\times10$~m evaluation area, considering $h_{u}=1.5$~m, $d_{\rm BR}=20$~m, $8\times8$ RIS, and $P_{max}=75$~dBm.}
  \label{e-field-2d}
 \end{figure*}
\subsection{E-field Expressions}
For the RO mode, the E-field can be obtained by using the absolute phase calculated geometrically. Considering the BO mode with $\phi_n-\phi_d=0$, we have the closed-form expression based on Eqs. (1),(2),(7),(8), which is written as
\begin{equation}
\label{ef_bo}
    E_u = \sqrt{60P_{max}}\frac{\lambda}{4\pi d_{\rm BR}}\left|\sum_{n=1}^{N^2}\frac{\sqrt{G_n(\theta_n)}}{r_n}\right|.
\end{equation}
To simplify the expression, we first rewrite $\theta_n$ as
\begin{equation}
\theta_n=\arctan\left(\frac{h_n-h_u}{\sqrt{r_n^2-(h_n-h_u)^2}}\right),
\end{equation}
where we denote $q_n=\frac{h_n-h_u}{\sqrt{r_n^2-(h_n-h_u)^2}}$ for simplicity.

Then, we have
\begin{equation}
\sqrt{G_n}=(\cos({\arctan{q_n}}))^{\frac{3}{2}}=\frac{1}{(1+q_n^2)^{\frac{3}{4}}}.
\end{equation}

Thus, Eq.~(\ref{ef_bo}) can be rewritten as
\begin{equation}
\begin{aligned}
E_u &= \sqrt{60P_{max}}\frac
  {\lambda}{4\pi d_{\rm BR}}\sum_{n=1}^{N^2}\frac{1}{(1+q_n^2)^{\frac{3}{4}} r_n},\\
&= \sqrt{60P_{max}}\frac{\lambda}{4\pi d_{\rm BR}}\sum_{n=1}^{N^2}\left(\frac{r_n^2-(h_n-h_u)^2}{r_n}\right)^{\frac{3}{4}},\\
\end{aligned}
\end{equation}
where we can obtain the E-field by using the coordinates.

\section{Evaluations}
 In this section, we first use Wireless InSite simulations to verify the E-field expression. Then, the peak E-field values are evaluated considering different RIS sizes and deployment configurations. Finally, various impacts of FR and patterns of elements are investigated.

\subsection{Wireless InSite Simulation}
We compare the model in Eq.~(9) with the results simulated by the Wireless InSite, a 3D ray-tracing simulator to analyze EM wave
propagation and wireless communication systems. The simulation objective is to verify the E-field based on Eq.~(\ref{eq:efield_user}) and Eq.~(\ref{eq:ele_field_f}), with a fixed $P_{RIS}$. Since the free-space path loss model is only dependent on frequency and distance, the elements in the RIS receive identical power from the BS. Therefore, we only analyze the E-field from the RIS to the receiving (Rx) points, without considering path loss between Primary Tx and RIS. We applied the X3D Ray Model propagation model, a full 3D propagation model achieved through the exact path calculation, and the ray spacing 0.25$^\circ$, enabling an accurate ray path between RIS elements and the receiving Rx points. We consider an open-space environment without any reflection from the other scatterers.

The RIS is represented as a $N\times N$ Uniform Rectangular Array (URA) with isotropic antenna elements spaced by half-wavelength between elements. We use a fixed 0~dBm total RIS power, representing the total power reflected by the RIS to the user. We evaluate the worst-case scenario, where the user or Rx points are located in the boresight direction of the RIS center, receiving the highest possible power from the RIS. The simulation outputs are $E_x$, $E_y$ and $E_z$, denoting the complex E-field in the Cartesian plane. The magnitude of the total E-field at the user is calculated by summing the complex-valued E-field contributed by all RIS elements as
\begin{equation}
    E_{u}=\sqrt{\left(\sum_{n=1}^{N^2} E_{x_n}\right)^2 + \left(\sum_{n=1}^{N^2} E_{y_n}\right)^2 + \left(\sum_{n=1}^{N^2} E_{z_n}\right)^2}.
\end{equation}
Note that the RIS works in the RO mode since there is no beamforming function in Wireless InSite simulations. Fig.~\ref{e-field-comparison} shows a comparison with good agreements between our analytical results and simulation results for different RIS  sizes, proving the correctness of the E-field expression. For instance, the peak values for $4\times 4$ RIS are 27.38~dBV/m and 27.61~dBV/m for the analysis and simulation, respectively.
 \begin{figure}[!t]
  \centering
   \subfigure[Fixed $h_{\rm RIS}=3$~m ]{\includegraphics[width=3in]{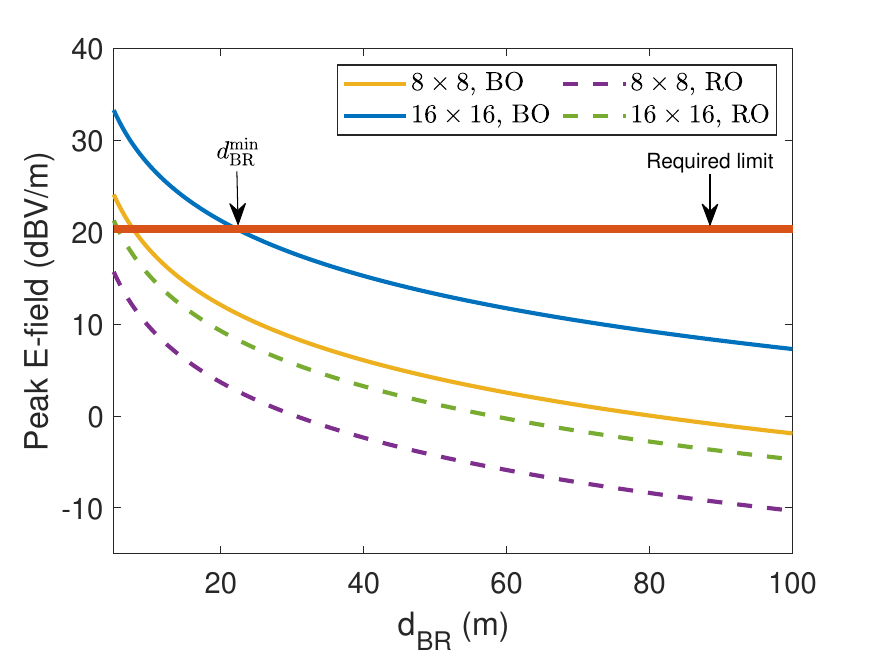}}
   \subfigure[Fixed $d_{\rm BR}=20$~m]{\includegraphics[width=3in]{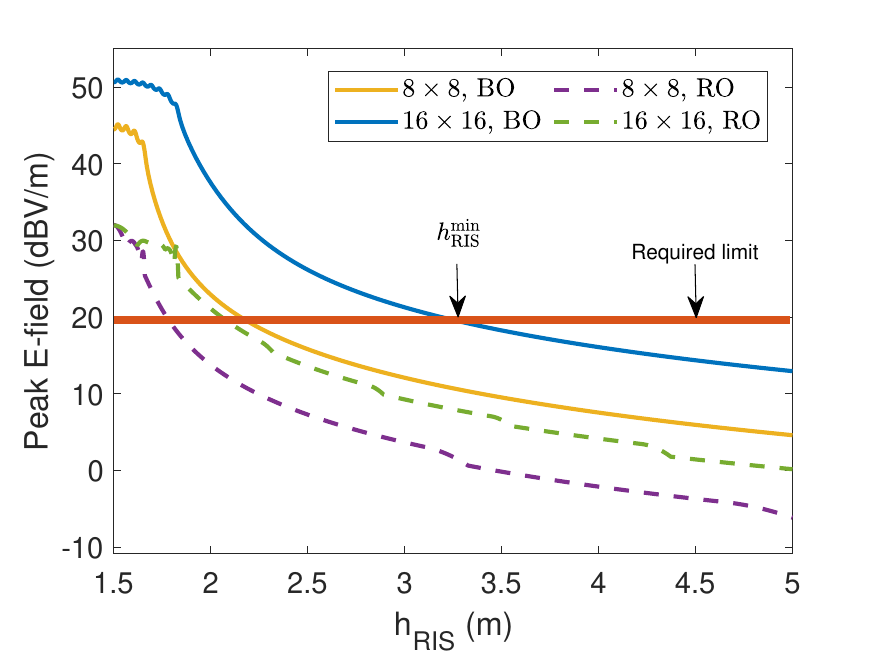}}
  \caption{Peak E-field as a function of $d_{\rm BR}$ and $h_{\rm RIS}$ for different $N_{\rm RIS}$, considering 75~dBm EIRP at BS and $G=\cos(\theta)^3$.} 
  \label{e-field-limit}
 \end{figure}

\subsection{2D E-field Map}
Considering a $10~m \times 10~m$ evaluation area, we can generate the E-field map. We consider two RIS placements at different heights, i.e., 1.5~m and 3~m. With $h_u=1.5$~m for practical consideration, Fig.~\ref{e-field-2d}(a) and (b) correspond that the RIS and users are located on the same level, where results from Wireless InSite and our model show perfect agreement, with only 0.2~dB difference regarding peak values. Fig.~\ref{e-field-2d}(c) and (d) suggest that the down-tilted beam from the RIS points to users, leading to grid coverage and beam spots when considering RO and BO modes respectively, where we set the target user coordinates on the $XOY$ plane as [0, 1]~m. By considering a single user for the BO mode, we create a worst-case beam spot scenario, with a maximal E-field. The result shows that the BO mode creates a significantly higher E-field. In general,
the low-height RIS placement, or the RIS at the same height as the user, is less practical because of the limited beamspotting gain for communication needs. 
When the RIS is placed above the user, the BO mode is of high interest since its E-field is always higher than that in the RO mode.

\subsection{Limit Analysis}
Following the E-field guidelines, we should guarantee that the peaks do not exceed the limit at any location in the considered zone. Thus, we apply a reference limit for illustration purposes, at 20~dBV/m, corresponding to 10~V/m. This is a relatively high limit, as the guidelines are cumulative and it is in general not possible to have such high values for a single deployment or frequency. As shown in Fig.~\ref{e-field-limit}(a), the peak E-field decreases with distance $d_{BR}$, considering a fixed $h_{\rm RIS}=3$~m and frequency $f=3.5$~GHz. For the considered limit, there is no distance requirement for the RIS to work in RO mode. However, with beamforming gain, the required minimum $d_{BR}$ becomes larger when employing a higher number of elements. It is observed that the RIS heights influence the E-field in Fig.~\ref{e-field-2d}. To this end, we evaluate the impact of height on the E-field to find the minimum $h_{\rm RIS}$. Fig.~\ref{e-field-limit}(b) draws the same conclusion as Fig.~\ref{e-field-limit}(a), and confirms that the required minimum $h_{\rm RIS}$ in the RO mode is lower than that in the BO mode. It shows the $h_{\rm RIS}^{\min}$ under $16\times 16$ RIS is 2.05~m and 3.19~m for the RO and BO mode, respectively. For practical indoor deployments, small height differences between the user and the RIS are needed. Thus, it is recommended that RO-RIS can be used indoors and BO-RIS can be used outdoors but deployed above 4~m, which guarantees the safety of EMF exposure.

\begin{figure}[!t]
  \centering
  \includegraphics[width=3in]{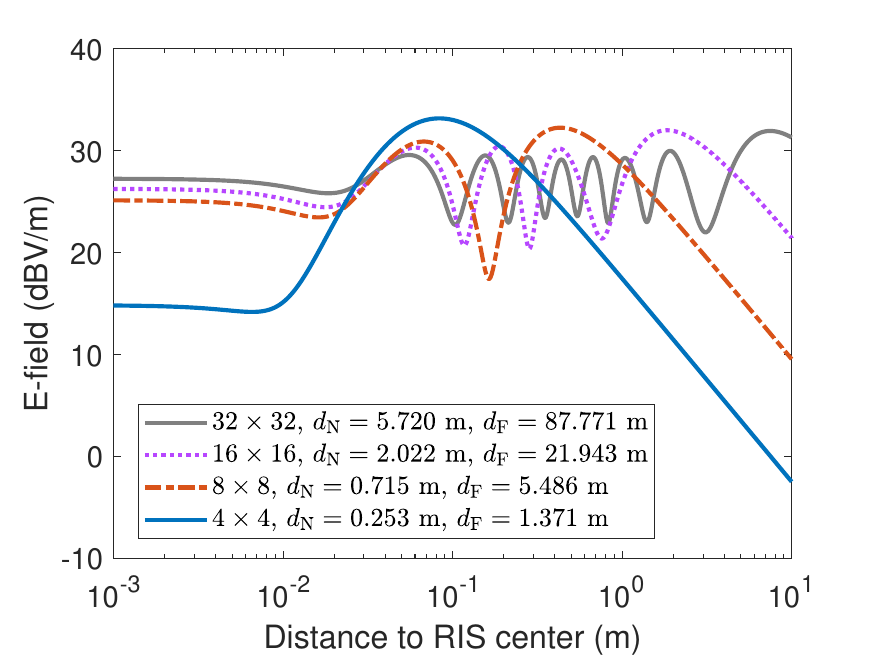}
  \caption{E-field in the RO mode in the bore-sight direction for different numbers of RIS elements where we consider $f=3.5$~GHz, $h_{\rm RIS}=1.5$~m, $d_{\rm BR}=20$~m, and $P_{max}=75$~dBm.}
  \label{near_peaks}
 \end{figure}

\subsection{E-field in RO mode}

Since the RO-mode RIS is energy-efficient as the need of power-hungry phase shifters is avoided, we also investigate the E-field in this setting. We are mainly interested in near-field scenarios, assuming an indoor RO-only RIS. The results are illustrated in Fig.~\ref{near_peaks}. First of all, it is observed that the number of peaks is determined by the RIS size. Moreover, from our simulations if follows that there are $\frac{N}{4}$ peaks. Close to the RIS, each location is only in field-of-view of a limited number of RIS elements, creating a first peak. As distance to the RIS increases, the number of RIS elements seen increases, and a second peak is created.  This also explains why the E-field is constant for very small distances, which corresponds to the original radiated field from a single element in the RIS. The final peak occurs because of the coherent superposition of all the spherical waves in the near-field region after the user can \textit{"see"} all elements. This finding can be used for studying the power distribution of near-field communications when employing large-size arrays and very short-distance scenarios. More importantly, our analysis shows that the maximum E-field values reside on the final peak when all elements contribute constructively to the signal received by the user. In the legend of the figure, we provide two distance indicators, one is denoted as the near-field distance that is the bound of the reactive region and Fresnel region of EM wave propagation, and the other one is the far-field distance that is the bound of the Fresnel region and far field, which are expressed as
\begin{equation}
    d_{\rm N}=0.62\sqrt{\frac{D_s^3}{\lambda}},\;\;\;
d_{\rm F}=\frac{2D_s^2}{\lambda},
\end{equation}
where $D_s$ is the largest dimension of the RIS, which can be expressed by $D_s=\frac{N\lambda}{\sqrt{2}}$ considering a URA and half-wavelength spaced elements. The results show that the peak values are 0.085~m, 0.432~m, 1.842~m, and 7.455~m approximately corresponding to $\lambda$, $5\lambda$ $21\lambda$, and $86\lambda$ for $N=4, 8, 16, 32$, respectively. It can be seen that the locations of peaks for $N=4, 8, 16$ are below both $d_N$ and $d_F$. However, the peak for a larger RIS of $N=32$ is between these two distances. Although the closed-form expression is intractable when considering the RO mode, the location of peaks can be obtained through the evaluation. In any deployment, the peak values should be suppressed below the limit, and the users are suggested to avoid the locations of peaks and reduce EMF exposure due to the time aggregation. Conversely, when the largest and furthest peak power is below the limit, all near-field locations are safe.



 \subsection{Impact of Element Pattern}
\begin{figure}[!t]
  \centering
   {\includegraphics[width=3in]{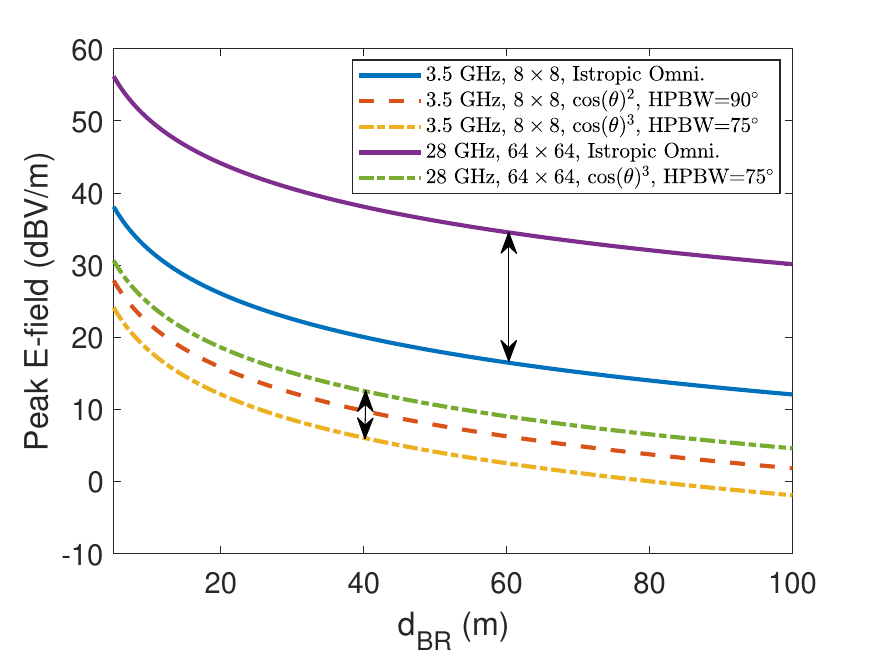}}
  \caption{Peak E-field in BO mode versus frequency and element pattern considering $h_{\rm RIS}=3$~m and $P_{max}=75$~dBm.}
  \label{E_field_pattern_frequency}
 \end{figure}

We consider different patterns of elements in the RIS, the peak E-field as a function of $d_{BR}$ is given in Fig.~\ref{E_field_pattern_frequency}. It shows that the high directivity, which is expressed by the narrow half-power beamwidth (HPBW), reduces the E-field. Moreover, when reducing the HPBW, the beamforming gain of high frequencies becomes smaller, see the differences indicated by two double arrows in the figure. In the evaluation, we consider the same RIS size for different frequencies, i.e., $8\times 8$ at 3.5 GHz corresponding to $64\times 64$ at 28 GHz. A more detailed evaluation in terms of FRs is given in the following subsection.

   \begin{figure}[!t]
  \centering
   {\includegraphics[width=3in]{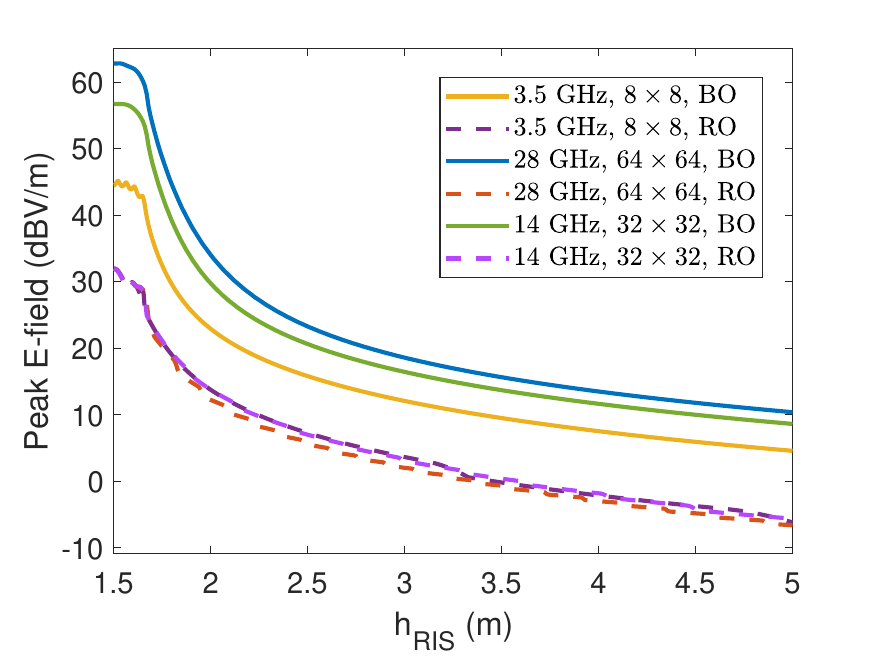}}
  \caption{Peak E-field versus frequency and RIS height considering $d_{\rm BR}=20$~m, $P_{max}=75$~dBm and $G=\cos(\theta)^3$.}
  \label{E_field_frequency_height}
 \end{figure}
\subsection{Impact of Frequency Range}
Considering the trend to use different FRs in future wireless networks, we investigate the E-field in FR1, FR2, and a potential new FR3 \cite{cui20236g} where the representative frequencies are 3.5~GHz, 28~GHz, and 14~GHz, respectively. Therefore, for a fair comparison, we employ the same RIS size, resulting in the numbers of elements being $8\times 8$, $64\times 64$, and $32\times 32$. Firstly, when working in the RO mode, the E-field values in the RIS-assisted networks are the same, which indicates the E-field is independent of the frequency, and determined by the power strength of the source (primary Tx). The observation also verifies that the RO mode of the RIS does not lead to a high risk of EMF exposure. The BO mode can result in risk when a using high amount of elements, since it is observed that the E-field increases with the number of RIS elements. 





\section{ Measurement Methodology}
\subsection{General Measurement Considerations in Cellular Networks}
Current regulations on EMF exposure focus on emissions from the BS rather than from the User Equipment (UE). Previous cellular systems use a Frequency Division Duplex (FDD) scheme, where part of the available spectrum is completely dedicated to uplink (UE to BS) and another part is dedicated to downlink (BS to UE).
In this system, measuring the emissions from BSs is a simple task, which consists of measuring power over the downlink channels. This is typically performed either with spectrum analyzers or dedicated EMF probes.
However, 5G networks completely changed the paradigm by employing Time Division Duplex (TDD) as well.
This means that the whole frequency band can be dynamically allocated either to downlink, uplink, or a combination of both. Moreover, some BSs can allocate part of the spectrum to older technologies such as LTE.
The uses of massive MIMO and RIS create a larger variation between the ratio of peak to average exposure in time, and between the users and non-users of the network at a given time.

\begin{figure}[t]
    \centering
    \includegraphics[width=0.88\linewidth]{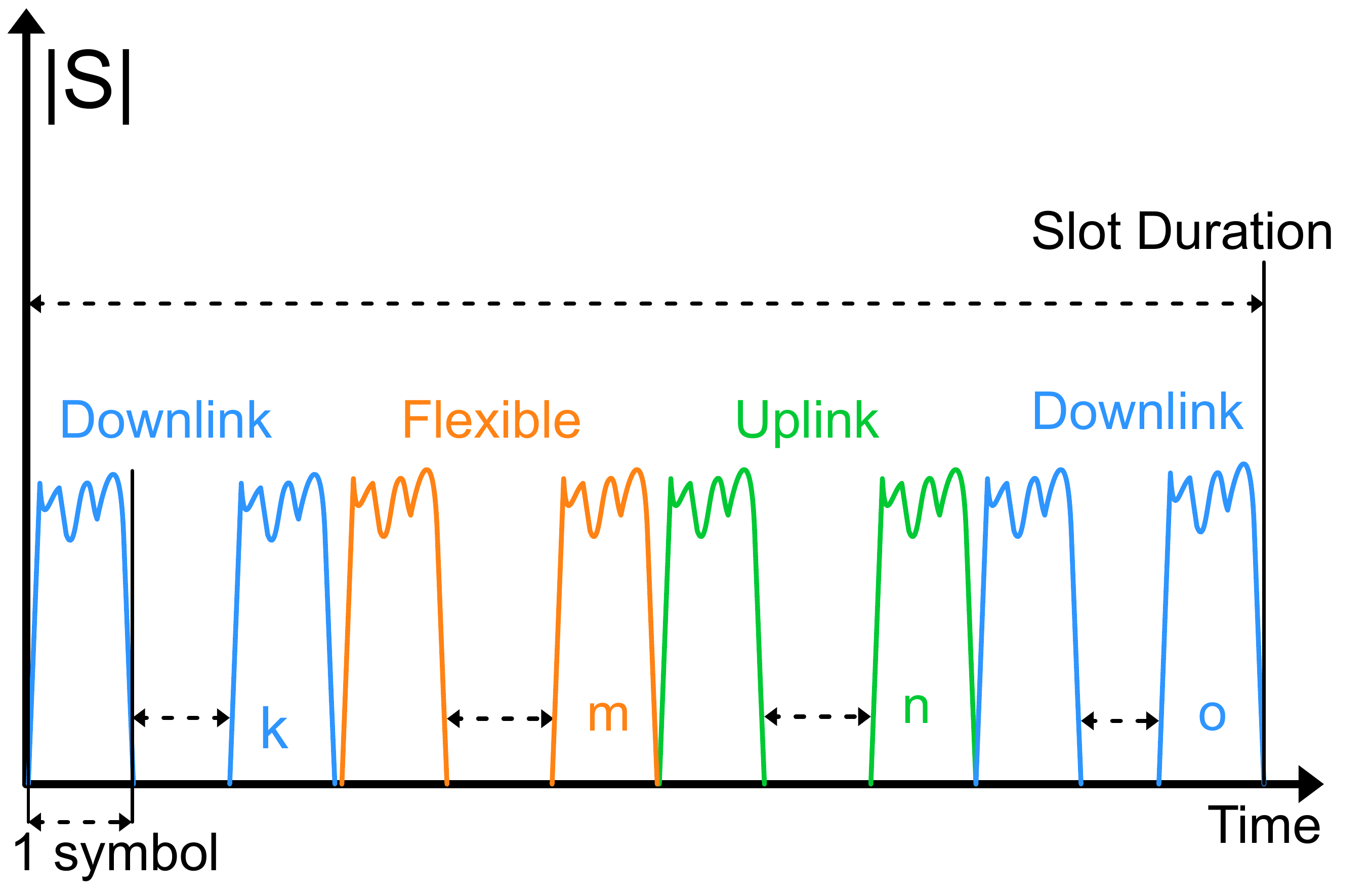}
    \caption{Schematic representation of a time slot. Each slot corresponds to 14 OFDM symbols ($k+m+n+o=14$ where $k,m,n,o$ are the number of symbols associated with each group). Some symbols are always allocated to downlink or uplink but the vast majority are "flexible".}
    \label{fig:frame_structure}
\end{figure}
The physical layer of 5G employs radio frames to implement TDD. Each radio frame is divided into subframes, and each subframe is divided into slots \cite{5g_frame_structure}.
Fig. \ref{fig:frame_structure} is a representation of a 5G slot. Each slot is composed of 14 OFDM symbols. While some symbols are always allocated to uplink or downlink, others can be dynamically allocated depending on user activity.
The slot duration, and thus the symbol time, can vary depending on the numerology of the BS and the available bandwidth. Because the time information is so important in determining how much energy is radiated by the user and how much by the BS, it is challenging for standard spectrum analyzers to distinguish between signals coming from the UE and signals coming from the BS.
The measurement procedure is simpler using instruments working in the time domain such as software-defined radios (SDRs) or real-time spectrum analyzers~\cite{5GFranco}. The EM field measurement details of 5G networks mentioned above also apply to RIS-assisted networks, which means the traditional measurement from primary Tx should be shifted to the measurement for the secondary Tx, i.e., RIS.

\begin{figure}[t]
    \centering
    \includegraphics[width=0.88\linewidth]{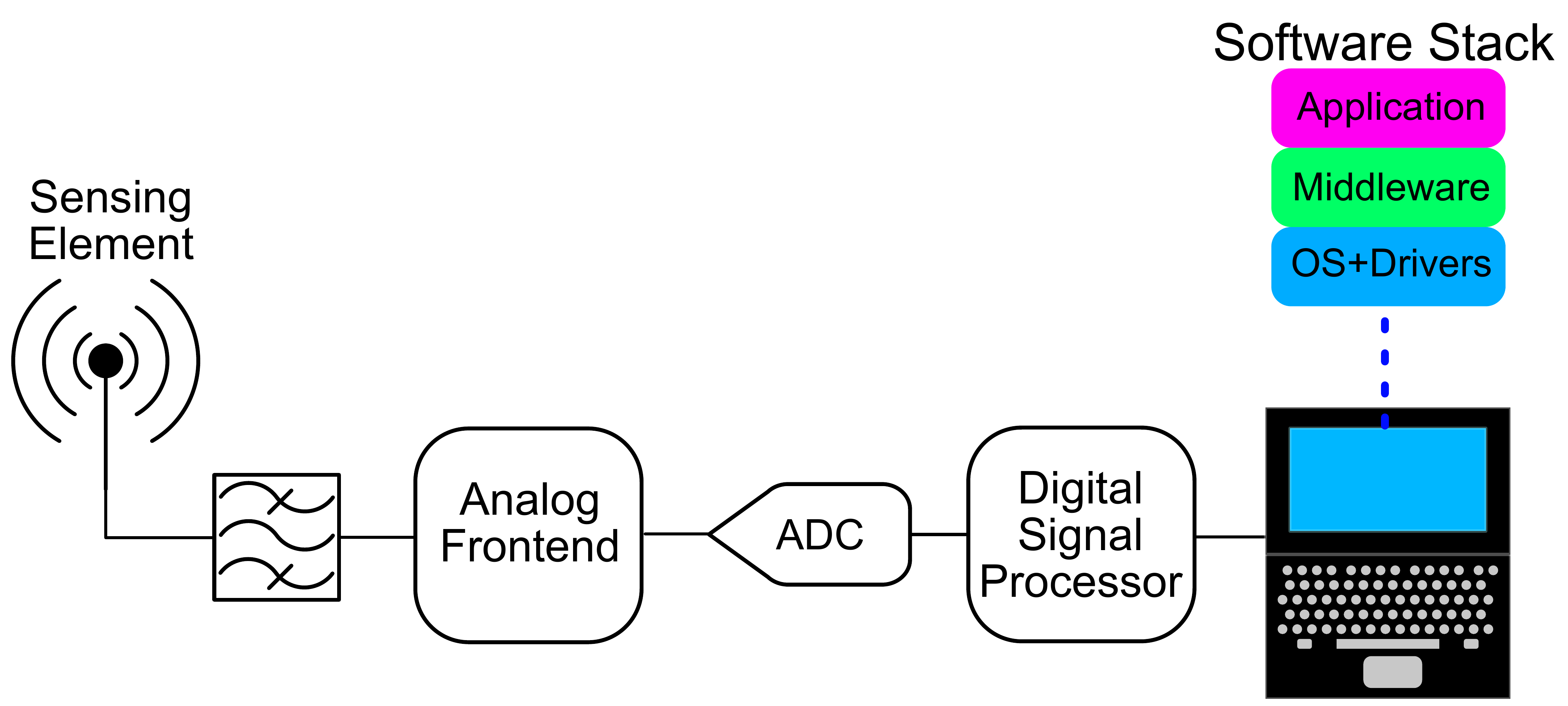}
    \caption{Schematic of a digital electric field sensor. The sensing element can be either an antenna or a field probe. The analog front end ensures that only the correct band is amplified, down-converted to the baseband, and passed on to the digital processing stage. The digital processing stage performs operations such as digital filtering and frame detection. Finally, a computer, which can be a laptop or an embedded PC, can post-process the data to calculate power, calculate the FFT, and manage the user interface.}
    \label{fig:td_sensor}
\end{figure}

Fig. \ref{fig:td_sensor} shows the typical sensor architecture used to measure electric fields from digital communication systems. This architecture can be implemented with an ad-hoc device \cite{narda_probe}, an SDR \cite{ReyndersBrecht2018UCRf,yassine,5GFranco}, or a real-time spectrum analyzer.
To get the correct value of received power and E-field, it is necessary to perform some digital processing on the received data.
The main tasks of the digital signal processing include the following aspects:
\begin{itemize}
    \item Find the beginning and end of each slot;
    \item Classify the symbols to detect which one comes from a base station and which one comes from the user;
    \item Calculate the power of each symbol over a specific amount of time;
    \item Calculate the equivalent electric field strength.
\end{itemize}
The time over which the power needs to be measured depends on the local legislation. It is based on an estimation of the time needed to raise the skin temperature of an individual by 1 \textcelsius~\cite{emf_guideline}. However, other considerations such as instrument limitations, cellular standards, or previous legislation, need to be considered when the regulator decides on the target measurement time.

\subsection{Recommendations for Measurement Procedures}

To monitor EMF exposure due to RIS, it is essential to perform many measurements for a long enough time. Governments, at the moment, can only perform spot measurements at a specific moment using expensive equipment.
However, with new technologies such as RIS, 5G, and the upcoming 6G networks, there will be more and more of a need for low-cost configurable nodes that can measure the E-field accurately over long times and at many locations. Thus, there is a strong need for low-cost, easy-to-deploy EMF sensors.

Special measurement procedures have been defined in 5G to measure peak spatial E-fields. Some proposed methods in the literature suggest forcing the beam towards the measuring device by generating a traffic demand from a nearby \textit{UE}.
This procedure is still not ideal as it is essential to accurately measure the field precisely at the \textit{UE}.
One of the many advantages of SDRs is that the device can mimic a user, thus forcing the BS to orient the beam directly toward the sensor. However, a mid-range SDR is required for this to work, such as a Universal Software Radio Peripheral (USRP).
While these radios are not the cheapest possible devices on the market, they are still less expensive than dedicated equipment like the one produced by Narda or Rohde\&Schwarz.
Especially in the case of RIS, where a wide area needs to be surveyed to find the highest E-field peak, deploying multiple devices at once can effectively save time and simplify the measurement procedure.
The signal can be measured considering the maximum capacity of the resource elements corresponding to the Physical Downlink Shared Channel (PDSCH)~\cite{5GdualTest}. This method has proven to accurately extrapolate to the maximum exposure level without prior knowledge of the radiation patterns~\cite{5GSwitzerland}.
The separation between uplink and downlink power can be performed post-processing by power discrimination and considering a certain ratio between uplink and downlink for the extrapolation over bandwidth and time~\cite{5GFranco,5GSwitzerland}. However, these methods do not consider additional spatial diversity~\cite{5g_radiated_SA} introduced by RIS, multi-band, and multi-user operations.


The measurements of the PDSCH should be complemented by extrapolating from measurements of the DL Physical Broadcast Channel (PBCH), especially when dealing with multi-bandwidth and carrier aggregation systems. This method also contributes to a better assessment of the ratio between mean and peak exposure.
Considering our results, we can conclude that when the RIS is used in RO, the best place to position EMF sensors is in front of the RIS, in the direction of reflection. On this line, moving away from the RIS, the E-field will show its peak values, which must remain below the allowed limits.
In the case of BO, the beam can be steered in different directions, thus moving the line where the E-field is the highest. Thus, placing the sensors in a semi-circle configuration in front of the RIS makes the most sense.
In both scenarios, there is still the need to "call the beam," such that the beam is oriented towards the sensing device.
To do that, we might envision two possible approaches. The first approach consists of permanently placing the sensors and letting users passing behind them call the beam. This approach allows statistics to be gathered over time, but it depends on how much traffic the users generate.
The second approach is temporarily deploying more sophisticated SDRs, which can imitate a user and thus call the beam themselves. This approach can lead to more predictable results, which can be gathered on command.

Depending on the positioning and structure of the RIS (i.e., altitude, indoor/outdoor, number of elements), E-field peaks might vary in distance and location from the center of the RIS. Through simulations, it is possible to determine both an optimal and a practical distance at which the sensors should be placed. In some instances, the peak E-field is just a few centimeters from the RIS (see the case of $4\times 4$ elements in Fig.~\ref{near_peaks}). Deploying a sensor at such a limited distance is not practical. Instead, a more pragmatic approach involves placing the sensor at the minimum accessible distance from the RIS.

For measuring the cumulative exposure over time for a multi-band and spatial diverse 5G-TDD network, we recommend investigating a distributed array of synchronized sensors capable of collecting data across multiple frequency bands in neighboring locations separated up to two times the angular beam size. This procedure is optional for certification of operator deployments but rather to collect data about the general public exposure over time and to detect anomalies (e.g., pirate transmitters or base stations not working correctly). This data can then be linked to health studies to help clarify whether long-term exposure affects the interested population.
For example, in \cite{extensive_geographical_monitoring}, the authors conducted E-field strength measurements in various locations by placing an RF probe on a car and driving in the area of interest.
To conclude, we suggest using receiver-only SDRs in the case of RO mode. We recommend active SDRs for BO mode, which can both call the beam and perform the EMF measurement.

\section{Conclusion}
In this paper, we focused on a RIS-assisted wireless network and looked into EMF exposure issues, by reviewing regulations and assessing E-field comprehensively. Firstly, both Wireless InSite simulation and the same E-fields using the same RIS size in the RO mode and different FRs prove the correctness of our analytical framework. Then, considering the passive RIS that works in the RO and BO modes, we found that the EMF exposure risk is very low in the RO mode, while the BO mode can lead to a high risk by using a high amount of elements. With the proposed analysis, the RIS deployment can be regulated by the minimum $d_{BR}$ and $h_{\rm RIS}$ to suppress the peak values below the safety guidelines. Moreover, various factors that can impact on E-field are analyzed, which shows that the high-directional pattern of the element helps reduce the beamforming gain, thus suppressing the above-limit E-field efficiently. To enable real-world E-field measurements, we finally introduce measurement devices, E-field calculation in practical 5G networks, and space-time-frequency measurements, which point out our future work.

\section*{Acknowledgement}
This work is supported by the MINTS project under the EU's Framework Programme for Research and Innovation Horizon 2020 with Grant Agreement No. 861222, and the 6G-Bricks project under the EU’s Horizon Europe Research and Innovation programme with Grant Agreement No. 101096954.

\bibliographystyle{IEEEtran}
\bibliography{Reference}


\end{document}